\documentclass[prl,twocolumn,superscriptaddress,showpacs,floatfix]{revtex4}
\usepackage{graphics}
\usepackage{graphicx}
\begin{document}

\title{Physical mechanisms of interface-mediated intervalley coupling in Si
}
\author{A.L. Saraiva}

\affiliation{Instituto de F\'{\i}sica, Universidade Federal do Rio de
Janeiro, Caixa Postal 68528, 21941-972 Rio de Janeiro, Brazil}

\author{M.J. Calder\'on}

\affiliation{Instituto de Ciencia de Materiales de Madrid (CSIC), Cantoblanco,
28049 Madrid, Spain}

\author{Xuedong Hu}

\affiliation{Department of Physics, University at Buffalo, SUNY, Buffalo, NY 14260-1500}

\author{S. Das Sarma}

\affiliation{Condensed Matter Theory Center, Department of Physics,
University of Maryland, College Park, MD 20742-4111}

\author{Belita Koiller}

\affiliation{Instituto de F\'{\i}sica, Universidade Federal do Rio de
Janeiro, Caixa Postal 68528, 21941-972 Rio de Janeiro, Brazil}

\date{\today}

\begin{abstract}

The conduction band degeneracy in Si is detrimental to quantum
computing based on spin qubits, for which a nondegenerate ground
orbital state is desirable.
This degeneracy is
lifted at an interface with an insulator, as the spatially abrupt change in the conduction band minimum
leads to intervalley scattering.
We present a theoretical study of the interface-induced valley
splitting in Si that provides simple criteria for optimal fabrication
parameters to maximize this splitting. Our work emphasizes the relevance of different interface-related properties
to the valley splitting.

\end{abstract}
\maketitle
\pagebreak
Semiconductor nanostructures based on GaAs and Si~\cite{hanson07,lansbergen08} are approaching the limit where device functionality relies on degrees of freedom of individual electrons. Recent progress in material processing allows precise controlled doping, band-structure engineering, and fabrication of high quality heterojunctions, which in turn pave the way for challenging applications such as the development of a scalable solid state quantum computer. The past few years witnessed tremendous experimental progress in the study of spin qubits at GaAs/AlGaAs quantum dots~\cite{hanson07}, which raises intriguing questions on the feasibility of spin qubits in Si quantum dot~\cite{friesen03} or donor states~\cite{calderonPRL06} at a Si/barrier-material interface.

A clear advantage of spin qubits in Si over GaAs is the long spin coherence times in Si~\cite{sarma04}. On the other hand bulk Si conduction band edge is six-fold degenerate, a complication not present in GaAs. Near a (001) interface with a barrier material, this degeneracy is partially lifted, with the interface electron ground state remaining doubly degenerate. For electron spin qubits, the  residual orbital degeneracy is an important spin decoherence source~\cite{tahan02}. This effect can be overcome if the ground state degeneracy is significantly lifted, which occurs close to an interface that can efficiently scatter carriers between the two degenerate valleys that are near opposite ends of the Brillouin zone \cite{ando82}. Measurements of the doublet splitting, or valley splitting, present significant variations among different Si/barrier samples, ranging from~$0$~to~$\sim$~1~ meV~\cite{goswamitakashina}. In this context, a simple physical model that can help identify the relevant fabrication-related parameters in order to maximize the valley splitting is a valuable tool in assisting current experimental efforts.

Theoretical approaches to describe the electronic behavior in the presence of an interface or heterojunction range from the effective mass approximation (EMA)  \cite{sham79,friesen07},  tight-binding models \cite{grossoboykin} to first-principles envelope function approach \cite{foreman05}. The present study, based on the physically motivated EMA~\cite{kohn}, aims to identify the relevance of sample-dependent parameters to the valley splitting. Our approach, while simple, is original and permits the study of the intervalley coupling due to a single interface, not employing periodic boundary conditions.
The full plane wave expansions of the Bloch functions at the two conduction band minima obtained from {\it ab initio} calculations \cite{koiller04} contain relevant physical information about the underlying Si substrate, and are included explicitly. We identify physical mechanisms that control the coupling strength pointing to convenient choices for the barrier material and the interface quality. We also highlight advantages and limitations of the EMA, in particular the oscillations of the valley coupling with the arbitrary choice of the interface position within a single monolayer distance~\cite{sham79}, which is an EMA artifact.

We consider the (001) Si/barrier system, so that the 6-valley Si bulk degeneracy breaks into a 2-valley ground state and a 4-valley excited state quartet $\sim 20$ to 30 meV above the ground doublet~\cite{kane00}. We write the Hamiltonian for this problem as
\begin{equation}
H = H_0 + U(z) - \frac{F}{\epsilon} z ,
\label{eq:single-hamilton}
\end{equation}
where $H_0$ is the unperturbed bulk Si Hamiltonian. Translational symmetry is assumed in the $xy$ plane, parallel to the interface, reducing the interface potential  to a $z$-dependent profile $U(z)$~\cite{bastard}. An electric field $F$ along the $z$ direction pushes the electron towards the interface. It is instructive to first consider an abrupt interface between Si and the barrier material; we model it here by taking $U(z)$ in Eq.~(\ref{eq:single-hamilton}) to be a simple step potential~\cite{bastard}
\begin{equation}
U(z) = U_{{\rm step}}(z) = U_0 \Theta (z-z_I),
\label{eq:step}
\end{equation}
with $U_0$ representing the barrier height. The usual barrier material in Si devices are SiO$_2$, corresponding to $U_0 \approx 3$ eV, and Si$_{1-x}$Ge$_x$ alloys, in which case $U_0$ may be tuned by controlling the alloy composition. A typical value, for $x\sim$ 0.2 to 0.3, would be $U_0 \approx 150$ meV.

\begin{figure}
\resizebox{75mm}{!}{\includegraphics{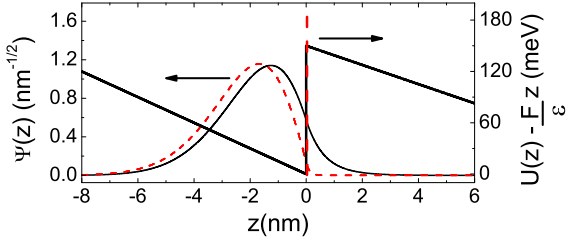}}
\caption{(Color online) Step model potential $U_{{\rm step}}(z)$ (thick lines) and
the calculated ground state envelope functions $\Psi(z)$ (thin lines) for the same
electric field $F=150$ kV/cm. The solid lines correspond to $U_0$ =
150 meV and dashed lines to $U_0=3$ eV, in which case the barrier
potential is above the vertical scale of the figure. The calculated valley splitting 
depends on $U_0$,
$|\Psi(0)|^2$ and the evanescent tail $\Psi(z>0)$.
}
\label{fig:profiles}
\end{figure}

Within single-valley EMA~\cite{kohn}, the lowest energy conduction band eigenfunctions for  Eq.~(\ref {eq:single-hamilton}) are written as
$\phi_\mu (\mathbf{r}) = \Psi(z) e^{i k_\mu z} u_\mu (\mathbf{r})$
where $\mathbf{k}_\mu= \pm k_0 \hat z$ are the Bloch wave vectors of the conduction band minima ($k_0\approx0.84\times 2\pi/a_0$). The envelope function $\Psi(z)$ satisfies the effective mass equation~\cite{bastard}
\begin{equation}
\left\lbrace \frac{-\hbar^2}{2 m_z} \frac{\partial^2}{\partial z^2} + U(z) - \frac{F}{\epsilon} z \right\rbrace \Psi(z) = E \Psi(z),
\label{eq:eff_mass_eq}
\end{equation}
where $m_z$ is the longitudinal effective mass for Si.
The ground state is numerically calculated through a finite differences method. Figure~\ref{fig:profiles} gives solutions for the above values of $U_0$. Strictly within the EMA assumption that the perturbation potential varies slowly in the length scale of the lattice parameter $a_0$, the ground state would remain doubly degenerate, since one obtains equivalent solutions for $\mathbf{k}_\mu= \pm k_0 \hat z$. This assumption is clearly not valid in the case of the step potential in Eq.~(\ref{eq:step}) which is discontinuous at the interface position $z=z_I$, thus coupling the originally degenerate Bloch valley states $|\phi_\pm (\mathbf{r})\rangle$. Also, the electric potential has a discontinuity in the derivative at $z=z_I$ due to the different values of the dielectric constant $\epsilon$ in the two materials. The interface position $z_I$ is initially taken to be $0$.
\begin{figure}
\resizebox{75mm}{!}{\includegraphics{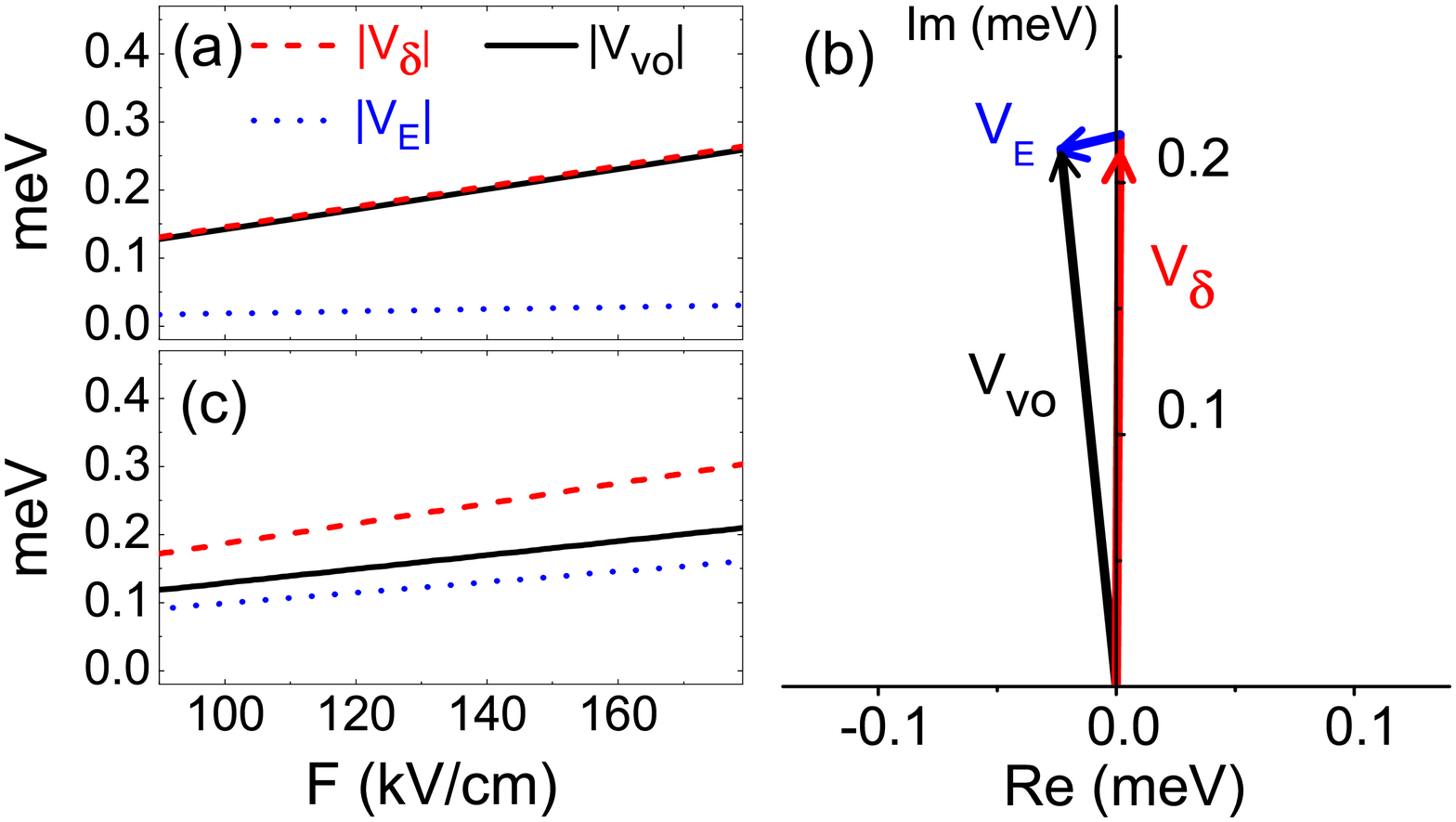}}
\caption{(Color online) (a) Absolute value of the intervalley coupling as a function of the electric field intensity for $U_0 = 150$ meV. The absolute values of the relevant terms in Eq.~(\ref{eq:decompose}) are also shown. (b) Representation of $V_{VO} = V_\delta + V_E$ in the complex plane for $U_0 = 150$ meV and $F=150$ kV/cm.  Although $V_E$ does not affect $|V_{VO}|$ much, it rotates $V_{VO}$ (i.e, changes its  phase). (c) Same as (a) for $U_0 = 3$ eV.}
\label{fig:terms}
\end{figure}

Since $\{|\phi_+\rangle,|\phi_- \rangle \}$ are well separated in energy from the excited states, an effective low energy Hamiltonian is
\begin{equation}
\bar{H} =\left( \begin{array}{cc}
 E_0 & V_{VO} \\
 V_{VO}^* &E_0
 \end{array} \right)~,
\label{eq:2by2matrix}
\end{equation}
where $E_0$ is the ground state energy obtained directly from Eq.~(\ref{eq:eff_mass_eq}), and the coupling due to the perturbation lifts the degeneracy leading to the valley splitting $2 |V_{VO}|$. The quantity of interest determining the splitting is $V_{VO}$, also called valley-orbit coupling, given here by
$V_{VO}=\langle \phi_+ | H | \phi_- \rangle$,
which is a complex number. Note that $H_0$ gives no contribution to this coupling. We write the periodic functions $u_\pm$ in terms of plane waves~\cite{koiller04}
\begin{equation}
\phi_\pm = \Psi(z) e^{\pm i k_0 z} \sum_{\mathbf G} c_\pm (\mathbf{G}) e^{i \mathbf{G} \cdot \mathbf{r}} ,
\label{eq:expansion}
\end{equation}
where  $\mathbf{G}$ are reciprocal lattice vectors. The expression for $V_{VO}$ then reads
\begin{equation}
V_{VO}=\sum_{\mathbf{G}, \mathbf{G^\prime}} c^*_+(\mathbf{G}) c_-(\mathbf{G^\prime}) \delta(G_x-G^\prime_x)\delta (G_y-G^\prime_y) I(G_z,G'_z),
\label{eq:sum}
\end{equation}
where the orthonormality of the $x$ and $y$ components is used, since there is no perturbation potential along these directions. The last term is an integral
\begin{equation}
I (G_z,G'_z)= \int_{-\infty}^{+\infty} |\Psi(z)|^2 e^{i Q z} \left[U_0 \Theta (z)- \frac{F}{\epsilon} z\right] \rm{d}z\\
\label{eq:integral}
\end{equation}
with $Q= G_z - G^\prime_z -2 k_0$. Terms with $\mathbf{G}$ and/or $\mathbf{G^\prime}\neq 0$ contribute to 
Eq.~(\ref{eq:sum}) with values  comparable to the $\mathbf{G} =\mathbf{G^\prime}= 0$ term alone. Integrating the contribution of the step potential  by parts, Eq. (\ref{eq:sum}) is rewritten as
\begin{equation}
V_{VO}= V_\delta + V_E + V_F~.
\label{eq:decompose}
\end{equation}
Here, $V_\delta$ is the contribution from the $I(G_z,G'_z)$ of the form  $\frac{iU_0}{Q} \int_{-\infty}^{\infty}e^{iQz}|\Psi(z)|^2\delta(z)\rm{d}z = i\frac{U_0}{Q}|\Psi(0)|^2$, similar to the effect of a $Q$-dependent $\delta$ function coupling potential. The contributions to $V_E$, of the form $\frac{iU_0}{Q} \int_{0}^{\infty}{e^{iQz}\frac{\rm{d}|\Psi(z)|^2}{{\rm d}z}\rm{d}z}$, involve only the evanescent part of the envelope function penetrating into the barrier region. We find  $V_F$, the electric field potential contribution, to be negligible in all cases considered here, meaning that the ``kink'' in the electric field potential at the interface does not couple $| \phi_\pm \rangle$, thus following the standard EMA assumption.
The role of the electric field in controlling $V_{VO}$ is mainly to modify the envelope function $\Psi(z)$.

Fig.~\ref{fig:terms}(a) shows the absolute value of the coupling, $|V_{VO}|$ and of the $|V_\delta|$ and  $|V_E|$ terms, for $U_0 = 150$ meV, as a function of applied field. In this case $|V_{VO}|$ is well described by the $\delta$-function contributions alone. If one is also interested in the eigenstates, the phase in $V_{VO}$ becomes relevant and the contribution $V_E$ plays a role, as illustrated in Fig.~\ref{fig:terms}(b). Fig.~\ref{fig:terms} (c) gives the results for the couplings dependence on the field for $U_0 = 3$ eV, and in this case both contributions affect $|V_{VO}|$. For the fields considered here, $|V_{VO}|$ increases linearly with $F$~\cite{sham79,ando82,grossoboykin}.
\begin{figure} [h!]
\resizebox{75mm}{!}{\includegraphics{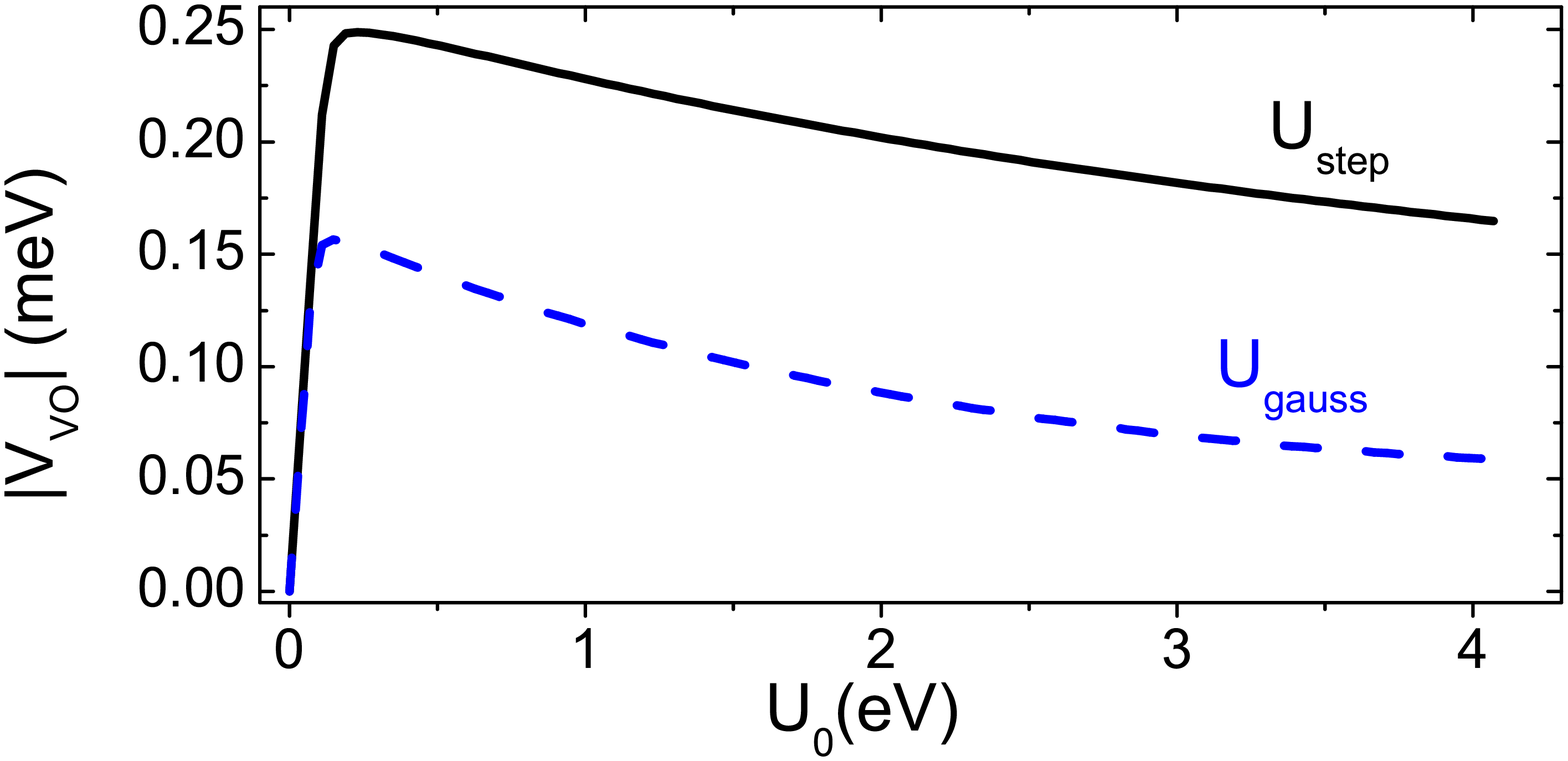}}
\caption{(Color online) Variation of the absolute value of the valley orbit coupling
with barrier height for $F=150$ kV/cm.  The step model of
Eq.~(\ref{eq:step}) and the smeared ($L=1$ monolayer) profile of Eq.~(\ref{eq:gauss})
are shown as the solid and dashed lines respectively. Both present a
maximum for $U_0 \approx 200$ meV.}
\label{fig:barrierheight}
\end{figure}

The expression for $V_{\delta}$ indicates that it is proportional to the product ${U_0}|\Psi(0)|^2$. As $U_0$ increases, $|\Psi(0)|^2$ decreases, so maximizing $|V_\delta|$ involves particular and not obvious conditions. The same is true for $V_E$, for which only the evanescent tail $(z>0)$ of
$\Psi(z)$ contributes. As shown in Fig.~\ref{fig:profiles}, when $U_0$ decreases the z-range that makes significant contribution to the integration increases.  However, the highly oscillatory phases $e^{iQz}$ in the integrand usually suppress this contribution.  A comparison of Fig.~\ref{fig:terms}(a) and (c) shows that the larger $U_0$ actually produces a much larger $|V_E|$, even though it corresponds to a more rapidly decaying evanescent envelope. Maximizing $|V_\delta|$ and $|V_E|$ independently does not necessarily maximize the absolute value of the complex sum $V_\delta + V_E = V_{VO}$. The net result is that $|V_{VO}|$ may be similar in magnitude even for materials with $U_0$ a factor of 20 apart, as shown in Fig.~\ref{fig:terms}. The general behavior of $|V_{VO}|$ with the barrier height $U_0$ is illustrated in  Fig.~\ref{fig:barrierheight} (solid line) for $F=150$ kV/cm. As $U_0$ increases from 0, a sharp rise in $|V_{VO}|$ is obtained up to a maximum coupling value around $U_0\approx 200$ meV, followed by a slow decay: Very low or very high barriers tend to suppress the valley coupling. Tuning $U_0$ involves changing the alloy composition in the barrier, as in the case of Si$_{1-x}$Ge$_x$, or changing the barrier material. Such fabrication processes are in principle experimentally feasible.
\begin{figure}[h!]
\resizebox{75mm}{!}{\includegraphics{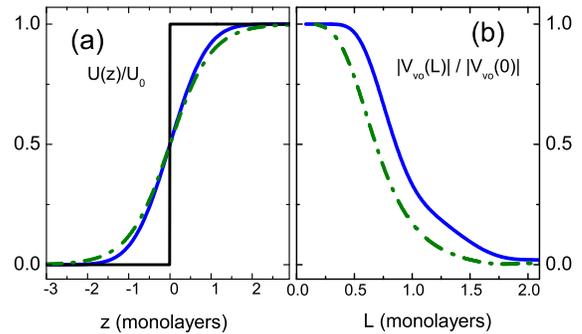}}
\caption{(Color online) (a) Normalized potential profiles for the
finite-width  interface models[Eqs.~(\ref{eq:exp}) and~(\ref{eq:gauss})]. The step potential is
also given. For $U_{{\rm gauss}}$ (solid curve)  and $U_{{\rm exp}}$
(dashed curve) the same width $L=1$ monolayer is taken.
(b) Variation with the interface width $L$ of the normalized valley
coupling for F=150kV/cm and $U_0=150$ meV  . The step potential
corresponds to $L=0$ in both models. Reported sample
measurements are consistent with $L \approx 0.5 - 1.5$ monolayers. Note that
the slightly steeper $U_{\rm gauss}$ significantly increases $|V_{VO}|$. The shape
of these curves is not strongly dependent on the value of U$_0$.}
\label{fig:VvsL}
\end{figure}

We now consider the effect of the interface width, which is disregarded in the step model.  We use for $U(z)$ in Eq.~(\ref{eq:single-hamilton} interface models that are similar to previously measured and calculated profiles \cite{sternishiyamabatsonyamasaki}), namely

\begin{eqnarray}
\label{eq:exp}
U_{{\rm exp}}(z) & = & \frac{U_0}{2} \left[\tanh({z/L})+1\right] ,\\
%
U_{{\rm gauss}}(z) & = & \frac{U_0}{2} {\rm Erfc} (-z/L) ,
\label{eq:gauss}
\end{eqnarray}
where ${\rm Erfc}(x)$ is the complimentary error function.
Both are characterized by a width $L$, and reproduce $U_{\rm step}(z)$ for $L=0$.
They differ in the asymptotic behavior (respectively exponential and gaussian). The curves are very similar [see Fig, 4(a) 
for $L = 1$  monolayer], as quantified by the RMS deviation with
respect to the step potential; the $U_{\rm exp}$ RMS is only 8$\%$ larger than that of the $U_{\rm gauss}$ profile. Yet, values of $|V_{VO}|$ differ by a factor of two. This factor may be as large as 3 (for L$\approx$1.3) with the same RMS ratio between the profiles. Fig.~\ref{fig:VvsL}(b) gives the calculated coupling versus $L$.

The interface width suppresses the intervalley coupling, as illustrated by the dashed lines in Fig.~\ref{fig:barrierheight}. The sensitivity of the coupling to the functional form of the interface potential for the same value of $L$ indicates that in real samples the coupling is strongly dependent on the type of interface disorder and roughness.  Both effects contribute to the experimentally observed variation in the intervalley scattering among different samples.

\begin{figure}[h!]
\resizebox{75mm}{!}{\includegraphics{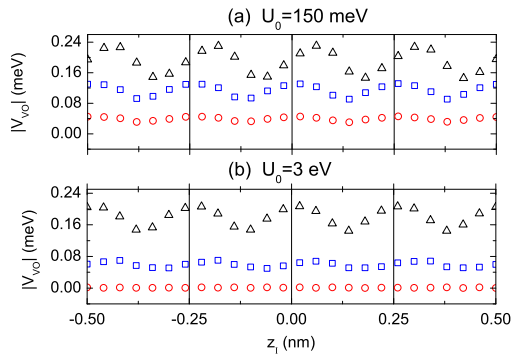}}
\caption{(Color online) Dependence of the intervalley coupling on the positioning of
the interface profile relative to the lattice atomic positions for
$U_0=$ (a)150 meV and (b)3 eV. The triangles represent  $U_{{\rm
step}}$ [Eq.(\ref{eq:step})], squares $U_{{\rm gauss}}$
[Eq.(\ref{eq:gauss})] and circles $U_{{\rm exp}}$
[Eq.(\ref{eq:exp})] , with  $L=$1 monolayer $\sim 0.13$ nm.}
\label{fig:Vvszi}
\end{figure}

A fundamental limitation of the step potential is the sensitivity of the results to the interface position $z_I$ (taken to be 0 in the results presented so far) on a sub-monolayer length scale.  This artifact, already obtained by Sham and Nakayama~\cite{sham79}, is due to the periodic parts of the Bloch functions, which carry information about the underlying Si lattice and lead to different interference patterns due to the interface perturbation according to its location. As expected, $|V_{VO}|$ shows oscillatory behavior with a 1 monolayer period, as seen in Fig.~\ref{fig:Vvszi}, but different models lead to
different relative phases. The more physical model with a finite interface width partially overcomes this artifact, as the amplitude of the oscillations is reduced for finite $L$. This is illustrated by the squares and circles in Fig.~\ref{fig:Vvszi}, and leads us to believe that a realistic model of the interface would largely remove this sensitivity to the exact interface location.

In summary, we have studied the conduction electron valley splitting in Si at a \emph{single} Si/barrier interface using EMA.  For a sharp interface, the simplicity of EMA allows us to identify two main contributions to the valley splitting: the electron wave function at the interface and the wave function gradient for the evanescent wave in the barrier material.  We show that an external field (to increase the wave function at the interface) and an appropriate barrier potential height (through proper choice of barrier material) can help maximize the valley splitting.

A shortcoming of the sharp interface model is that its results are highly sensitive to the interface position. Taking into account the finite interface width reduces the sensitivity to the interface location while also reducing the intervalley scattering, because it blunts the sharp intervalley interference.  This is illustrated by our result that steeper interfaces always favor larger intervalley splitting, while smoother profiles tend to reduce it, and may even lead to negligible coupling, as demonstrated by the $U_{\rm exp}$ profile in Fig.~\ref{fig:Vvszi}(b).

We do not expect our EMA results to be quantitatively accurate.  Many effects are not included, such as strain, interface misorientation~\cite{friesen07}, atomic scale disorder, lateral confinement, or many-body corrections to the valley splitting.  Nonetheless, the splittings $2|V_{VO}|$ on the order of 0.5 meV we obtained are in fair agreement with available measurements in Si/SiO$_2$ and Si/SiGe interfaces \cite{goswamitakashina}.

In conclusion, we have calculated electron valley splitting in Si at a Si/barrier interface.  We show that a sizeable single-particle splitting can be generated by applying a proper external field, choosing an optimal barrier material of a suitable potential height and producing sharp interfaces.  Lateral confinement and many-body corrections can potentially further increase this splitting.

\begin{acknowledgments}
We thank Rodrigo Capaz for helpful and fruitful discussions. This work was partially supported by the Brazilian agencies CNPq, FUJB, Millenium Institute on Nanotechnology - MCT, and FAPERJ. MJC acknowledges MAT2006-03741 and the Ram\'{o}n y Cajal program  (MICINN, Spain). XH and SDS thank financial support by NSA and LPS through US ARO, and by NSF.
\end{acknowledgments}
\bibliography{saraiva_short}
\end{document}